\documentclass[fleqn]{elsart}

\usepackage{amsmath}
\usepackage{amsfonts}
\usepackage{graphicx}
\usepackage{bm}

\def\be{\begin{equation}}
\def\ee{\end{equation}}
\def\ba{\begin{eqnarray}}
\def\ea{\end{eqnarray}}
\def\bs{\begin{subequations}}
\def\es{\end{subequations}}

\def\p{\partial}
\def\cS{{\cal S}}
\def\cL{{\cal L}}
\def\tphi{\tilde\phi}
\def\B{\Box}
\def\a{\alpha}

\def\s{\sigma}
\newcommand{\Eq}[1]{(\ref{#1})}

\begin{document}

\begin{frontmatter}

\title{Localization of nonlocal theories}

\author{Gianluca Calcagni},
\ead{g.calcagni@sussex.ac.uk}
\address{Department of Physics, University of Sussex, Brighton BN1 9QH, United Kingdom}
\author{Michele Montobbio},
\ead{montobb@science.unitn.it}
\address{Dipartimento di Fisica and INFN Gruppo Collegato di Trento, Universit\`a di Trento, via Sommarive 14, 38100 Povo (Trento), Italia}
\author{Giuseppe Nardelli},
\ead{nardelli@dmf.unicatt.it}
\address{Dipartimento di Matematica e Fisica, Universit\`a Cattolica,
via Musei 41, 25121 Brescia, Italia}
\address{INFN Gruppo Collegato di Trento, Universit\`a di Trento, 38100 Povo (Trento), Italia}
\date{\today}

\begin{abstract}
We show that a certain class of nonlocal scalar models, with a kinetic operator inspired by string field theory, is equivalent to a system which is local in the coordinates but nonlocal in an auxiliary evolution variable. This system admits both Lagrangian and Hamiltonian formulations, and its Cauchy problem and quantization are well-defined. We classify exact nonperturbative solutions of the localized model which can be found via the diffusion equation governing the fields.
\end{abstract}



\end{frontmatter}


\section{Introduction}

Nonlocal theories defined by pseudo-differential equations have recently attracted the attention of several research groups. The effective action for the tachyon field of open string field theory (OSFT) contains an infinite number of derivatives and falls into this class of models. As their dynamics is quite different with respect to the local one \cite{EW}, the search for solutions turned out to be a difficult task (\cite{MZ,FGN,CST,FGN2,KORZ,roll} and references therein). Minkowski kink-type solutions interpolating between different vacua and supposed to describe $D$-brane decays have been found numerically for the $3$-adic \cite{MZ,vol03} and supersymmetric string \cite{vol03,AJK}, while existence theorems were given in \cite{vla05,pro06}. The string tachyon has been also embedded in a cosmological context \cite{are04,AJ,cutac,AK,AV,BBC,kos07,AKV,AV2,lid07,BC,cuta2,jou07}, while nonlocality has found other applications in cosmology \cite{SW,BMS,kho06,DW}. In parallel, much effort has been devoted to the understanding of the mathematical structure of nonlocal theories \cite{vla05,pro06,VV,vla07}, especially as regards the initial conditions problem \cite{MZ,cutac,cuta2,BK}. For other applications in physics and an overview of the literature, see \cite{BK}.

The Cauchy problem is an example of the difficulties one meets when dealing with nonlocal kinetic operators. Apparently one should specify an infinite number of initial conditions; this would correspond to know the Taylor expansion of the solution around the initial point \cite{MZ}, in contrast with the usual physical interpretation. In the context of cosmological inflation, the infinite number of degrees of freedom leads to an involved definition of the slow-roll approximation \cite{cutac}. Only very recently, however, it was argued that the number of nonlocal degrees of freedom is actually finite \cite{cuta2,BK}. Moreover, the construction of nonperturbative solutions is a highly nontrivial task also on a Minkowski metric. While numerical methods can be of help in this respect \cite{vol03,jou07}, it would be desirable to have analytic tools at hand; one of these, based on the heat equation \cite{FGN,vla05}, proved to be promising \cite{roll,cuta2,jou07}, but its mathematical foundations were still incomplete.

In this Letter we want to address these issues and show how a wide class of nonlocal theories, including bosonic cubic SFT, can be thought as local systems living in a higher-dimensional space. After an introduction on the nonlocal model, the Lagrangian and Hamiltonian formulations of the \emph{localized} system are presented. Immediate byproducts of this construction are: (1) The diffusion equation \emph{ansatz} employed in \cite{roll,cuta2} in order to find nonlocal solutions is a key ingredient for the recovery of the equations of motion and stress-energy tensor; (2) The nonperturbative Cauchy problem is set up by a finite number of initial conditions (namely, four); (3) This is also the number of Hamiltonian degrees of freedom, resulting in a well-defined quantization of the fields. Exact nonperturbative solutions are classified and nontrivial examples are given.


\section{Nonlocal system}

We consider a nonlocal matter action of the type
\be
\cS_{\phi} = \int \d^D x \sqrt{-g}\,\cL_\phi,
\ee
where
\be\label{nlocph}
\cL_\phi =\tfrac12\phi\B\phi-U(\tphi),
\ee
and
\be
\tphi\equiv \e^{r_*\B}\phi\,.
\ee
Here $r_*$ is a parameter eventually fixed by the theory ($r_*\equiv \ln 3^{3/2}-\ln 4\approx 0.2616$ in string theory), $g$ and $\B$ are the determinant and d'Alembertian for a general $D$-dimensional  metric with signature $({-}{+}{+}\cdots{+})$, and we have neglected a local mass term without loss of generality. The `kinetic potential' $U$ is a function of the field $\tphi$ with no other insertions of nonlocal operators $\e^{r_*\B}$. This case encompasses the bosonic string, $U\propto \tphi^3$ \cite{KS2}, and a simplified version of the supersymmetric string, $U\approx \tphi^4$. The more realistic susy case, $U\propto (\e^{r_*\B}\tphi^2)^2$ \cite{AJK}, is not contemplated for technical reasons soon to be clear.

Applying the variational principle, the scalar equation of motion is
\be\label{pheom}
\B\phi-\e^{r_*\B} U'=0\,,
\ee
where a prime denotes derivative with respect to $\tphi$. For later convenience we recast this equation as
\be\label{tpheom}
\B \e^{-2r_*\B}\tphi-U'=0,
\ee
which can be derived from the Lagrangian density 
\be\label{nloctph}
\cL_{\tphi} =\tfrac12\tphi\B \e^{-2r_*\B}\tphi-U(\tphi)\,.
\ee
Nonlocal operators now appear only in the kinetic term, while the potential is local. 

The covariant stress-energy tensor of the scalar field is
\be\label{dt}
T_{\mu\nu}^{(\tphi)}\equiv -\frac{2}{\sqrt{-g}}\frac{\delta \cS_{\tphi}}{\delta g^{\mu\nu}}\,,
\ee
where the indices $\mu$ and $\nu$ run from 0 to $D-1$ dimensions and the action $\cS_{\tphi}$ is defined via Eq.~\Eq{nloctph}. Then, using the same lengthy procedure as in \cite{AJK,cutac,yan02},
\be\label{tmunu}
T_{\mu\nu}^{(\tphi)} =T_{\mu\nu}+ \int_0^{r_*} \d s\, \tilde T_{\mu\nu}^{(\tphi)}(s),
\ee
where 
\be\label{tloc}
T_{\mu\nu}\equiv \p_\mu\tphi\p_\nu \e^{-2r_*\B}\tphi-g_{\mu\nu}\left(\tfrac12\p_\s\tphi\p^\s \e^{-2r_*\B}\tphi+U\right)\,,
\ee
and
\ba
\tilde T_{\mu\nu}^{(\tilde\phi)}(s)&\equiv& g_{\mu\nu}\left[(\B \e^{2(s-r_*)\B}\tphi)(\B \e^{-2s\B}\tphi)+(\p_\s \e^{2(s-r_*)\B}\tphi)(\p^\s \B \e^{-2s\B}\tphi)\right]\nonumber\\
&&\qquad-2(\p_\mu \e^{2(s-r_*)\B}\tphi)(\p_\nu\B \e^{-2s\B}\tphi).
\ea
So far the scalar field has been considered as a function of the coordinates, $\tphi=\tphi(x)$. In order to find nonperturbative (in $r_*$) solutions of this system, a method was devised which localizes the equations in $x$ by promoting the constant $r_*$ to an evolution parameter $r$, and the scalar field as one living in $1+D$ dimensions, $\Phi(r,x)$, and such that
\be
\tphi(x)=\Phi(r_*,x)\,.
\ee
Imposing the \emph{diffusion equation} \cite{roll,vla05,cuta2}
\be\label{tdif}
(\B-\gamma\p_r-\beta)\Phi(r,x)=0\,,
\ee
nonlocal operators act as translations in the auxiliary variable $r$, so that 
\be\label{trasl}
\e^{qr(\B-\beta)}\Phi(r,x)= \Phi(r+\gamma qr,x).
\ee
The shifted function obeys a diffusion equation $(\B-\a\p_r-\beta)\Phi(r+\gamma qr,x)=0$ with $\a=\gamma/(1+\gamma q)$.
Nonlocal solutions can be constructed starting from a solution of the local system ($r=0$) which is interpreted as the initial condition $\Phi(0,x)$ of a system evolving in $r$ according to Eq.~\Eq{tdif}, such that $\Phi(r,x)= \e^{r(\B-\beta)/\gamma}\Phi(0,x)$. The constant coefficients $\gamma$ and $\beta$ are chosen accordingly.\footnote{In the conventions of \cite{vla05,jou07} $\beta=0$ and the initial condition at $r=0$ is not the local solution of the local system but the nonlocal field itself, $\Phi(0,x)=\tilde\phi(x)$. The constant $\beta$ can be set to zero whenever the normalization of the potential is thought to be $r$-independent; this was not the attitude of \cite{FGN,roll,cuta2}.}

With the \emph{ansatz} \Eq{tdif}, it was possible to find approximate solutions for the bosonic and susy string tachyon \cite{FGN,roll} and for cosmological toy models \cite{cuta2}. However, two problems remained unsolved: Are all solutions of the system Eq.~\Eq{nloctph} governed by the diffusion equation? Do nontrivial exact solutions exist for a given potential?

Here we want to provide evidence in support of affirmative answers. The first step is to incorporate the diffusion equation in a system which is $(1+D)$-dimensional and \emph{local} in $D$ dimensions, and show that this system not only is equivalent to the nonlocal one with the ad-hoc Eq.~\Eq{tdif}, but also naturally reproduces the structure of the nonlocal stress-energy tensor.


\section{Localized diffusing system}

We begin with the localized matter action in $1+D$ dimensions
\be\label{act}
\cS_{\Phi}=\int \d^D x \sqrt{-g}\int \d r \left(\cL_{\Phi}+\cL_{\chi}\right)\,,
\ee
where the two Lagrangian densities read
\be\label{locPh}
\cL_{\Phi} =\tfrac12\Phi(r,x)\B \Phi(r-2\gamma r,x)-V[\Phi(r,x)]\,,
\ee
and\footnote{Diffusing systems of scalar fields are discussed, e.g., in \cite{MF}. The noticeable difference between them and Eq.~\Eq{locch} is an extra integration.}
\be\label{locch}
\cL_{\chi} = \int_0^r \d s\,\chi(r',x)(\gamma\p_{r''}-\B+\beta)\Phi(r'',x)\,.
\ee
Here, $g$ is the determinant of the $D$-dimensional metric which does not depend on $r$, $\Phi$ and $\chi$ are $(1+D)$-dimensional scalar fields, $V\equiv \e^{2\beta r}U$, $\gamma$ and $\beta$ are constants, and 
\be
r' \equiv r-2\gamma s\,,\qquad r''\equiv r(1-2\gamma)+2\gamma s\,,
\ee
are linear combinations of $r$ and $s$. The action is therefore local on the submanifold with metric $g_{\mu\nu}$ and nonlocal in $r$. The $r$-integral in Eq.~\Eq{act} runs from 0 to an arbitrary upper bound (even $\infty$), whose value does not affect what follows. The lower bound is dictated by the initial condition in $r$, the latter being the local solution ($r=0$) from which the localized solutions are constructed as explained in \cite{roll,cuta2}.

We have calculated the variations of the action by using the functional derivative $\delta f(r,x)/\delta f(\bar r, \bar x)=\delta(r-\bar r)\delta^{(D)}(x-\bar x)$ for a field $f$. Due to the arbitrariness of $\bar x$ and $\bar r$ and requiring the support of the $\delta$'s to lie within the integration domains, variation with respect to the field $\chi(\bar r,\bar x)$ yields Eq.~\Eq{tdif}, while $\delta\cS_\Phi/\delta\Phi(\bar r,\bar x)=0$ if two equations are simultaneously satisfied:
\be\label{chiphi}
\chi(r,x) = \B\Phi(r,x)\,,
\ee
and
\be\label{locpheom}
\B \Phi(r-2\gamma r,x)-V'[\Phi(r,x)]=0.
\ee
In particular, $\chi$ satisfies the same diffusion equation for $\Phi$. Equation \Eq{locpheom} is in agreement with the original scalar equation \Eq{tpheom} if $r=r_*$.

The Hamiltonian and initial conditions problem are in general ill-defined in nonlocal theories, as the equations of motion contain infinitely many derivatives to be specified. An important feature which now emerges is that the Cauchy problem of the localized system is well-posed, as there is a finite number of initial conditions to be specified. Since the equations of motion for the two fields are second-order, the initial conditions are $\Phi$, $\dot\Phi$, $\chi$, $\dot\chi$, evaluated at a sensible initial point $x_0$. Because of Eq.~\Eq{chiphi}, they correspond to values of $\Phi$ and its first three derivatives. Such a result is in agreement with \cite{BK}, where it was argued that the number of initial conditions of the free system ($V\sim\Phi^n$, $0\leq n\leq 2$) is the number of poles in the propagator. In our class of nonlocal Lagrangians, the propagator accounts for two degrees of freedom; the other two are nonperturbative and come from a general potential (nonlinear equation of motion), via the field $\chi$. Intuitively, the infinite number of degrees of freedom of the nonlocal system have been transferred to the initial condition $\Phi(0,x)$ for the diffusion equation, which is a field in the coordinates $x$ evaluated at the particular value $r=0$ \cite{roll,cuta2,MF}.

As the fields are considered at different values of the extra `radial' coordinate, the action is not covariant in $1+D$ dimensions, as is also clear looking at $\cL_\chi$. However, it is covariant in the metric of the $D$-dimensional submanifold, and it makes sense to define a 2-tensor $T_{\mu\nu}^{(\Phi)}$ as in Eq.~\Eq{dt} with $\cS_{\tphi}\to \cS_{\Phi}$ and the indices still running on the $D$-manifold coordinates. In order to complete the proof of equivalence between the nonlocal and localized systems we show that
\be
T_{\mu\nu}^{(\Phi)}=\e^{2\beta r} T_{\mu\nu}^{(\tphi)}\,.
\ee
It is easy to see that\footnote{Strictly speaking, the metric is first thought as a function of all coordinates, $g_{\mu\nu}=g_{\mu\nu}(r,x)$; after functional differentiation of the action, all terms in $\p_r g_{\mu\nu}$ are dropped.}
\be\label{1tmunu}
T_{\mu\nu}^{(\Phi)} =T_{\mu\nu}+\int_0^r \d s\, \tilde T_{\mu\nu}^{(\Phi)}(s)\,,
\ee
where $T_{\mu\nu}$ reproduces Eq.~\Eq{tloc},
\be
T_{\mu\nu} =\p_\mu\Phi(r,x)\p_\nu \Phi(r-2\gamma r,x)-g_{\mu\nu}\left[\tfrac12\p_\s\Phi(r,x)\p^\s\Phi(r-2\gamma r,x)+V\right],
\ee
and
\be
\tilde T_{\mu\nu}^{(\Phi)}(s)\equiv g_{\mu\nu}(\p_\s\chi\p^\s\Phi+\chi\B\Phi)-2\p_\mu\chi\p_\nu\Phi,
\ee
the arguments of $\chi$ and $\Phi$ being $r'$ and $r''$, respectively. Using Eqs.~\Eq{trasl} and \Eq{chiphi}, the matching is complete. In the localized system the calculation of the energy tensor is straightforward, since one has no longer to deal with the variation $\delta \e^{r\B}/\delta g^{\mu\nu}$.


\section{Hamiltonian formalism}

The localized system admits a Hamiltonian formulation. In fact, there is a finite number of conjugate momenta. Defining the matter Lagrangian
\be
L\equiv \int \d^{D-1}x\int \d r \sqrt{h}\,(\cL_\Phi+\cL_\chi),
\ee
where $h$ is the determinant of the spatial $(D-1)$-metric, one has (we set the lapse function to unity and the shift function to 0)
\ba
\pi_\Phi(\bar r,x) \equiv \frac{\delta L}{\delta\dot\Phi(\bar r,x)}
                    &=&\frac{\sqrt{h}}{2}\left[\dot\Phi(\bar r(1-2\gamma),x)+\frac{1}{|1-2\gamma|}\dot\Phi\left(\frac{\bar r}{1-2\gamma},x\right)\right.\nonumber\\
                   &&\left.-\frac{1}{|\gamma|}\int \d r\, \dot\chi(2r(1-\gamma)-\bar r,x)\right]\,,\label{pif}\\
\pi_\chi(\bar r,x) \equiv \frac{\delta L}{\delta\dot\chi(\bar r,x)}
                    &=&-\frac{\sqrt{h}}{2|\gamma|}\int \d r\, \dot\Phi(2r(1-\gamma)-\bar r,x)\,.\label{pic}
\ea
The nonvanishing equal-time Poisson brackets are
\ba
\{\Phi(r_1,x_1),\,\pi_\Phi(r_2,x_2)\}_{t_1=t_2} &=& \delta(r_1-r_2)\,\delta^{(D-1)}({\bf x}_1-{\bf x}_2)\,,\\
\{\chi(r_1,x_1),\,\pi_\chi(r_2,x_2)\}_{t_1=t_2} &=& \delta(r_1-r_2)\,\delta^{(D-1)}({\bf x}_1-{\bf x}_2)\,,
\ea
where ${\bf x}_i$ are spatial $(D-1)$-vectors. The matter Hamiltonian reads
\be
H=\int \d^{D-1} x\int \d r \left[\pi_\Phi(r,x)\dot\Phi(r,x)+\pi_\chi(r,x)\dot\chi(r,x)- \sqrt{h}(\cL_\Phi+\cL_\chi)\right]\,,
\ee
where the total Lagrangian is written in terms of first derivatives of the fields. We have checked that the equations of motion are recovered. The evolution equations for the fields $\Phi$ and $\chi$ reproduce the momenta Eqs.~\Eq{pif} and \Eq{pic}, while
\be
\dot\pi_\chi( r,x)=\{\pi_\chi( r,x),\,H\}
\ee
gives the diffusion equation for $\Phi$, Eq.~\Eq{tdif}. The other Hamilton equation
\be
\dot\pi_\Phi( r,x)=\{\pi_\Phi( r,x),\,H\}\,,
\ee
yields the equation of motion for $\Phi$, Eq.~\Eq{locpheom}, provided Eq.~\Eq{chiphi} holds. Due to the complicated structure of the arguments of the fields, these results are nontrivial as they rely upon a number of delicate cancellations.

Once the classical Hamiltonian system is consistently constructed, one may be interested in the quantization of the fields, the Poisson brackets being replaced by commutators. Here we have shown that quantization is well-defined as the number of degrees of freedom is finite, which is not apparent in the original nonlocal Lagrangian system. 

There are no ready applications of this result within string field theory, as the tachyon action is already effective. However, our approach is helpful to quantize other nonlocal models, as is the case of cosmological perturbations of a nonlocal scalar field.

In \cite{cutac,cuta2} it was suggested a similarity between the diffusion equation method and the $1+1$ constrained Hamiltonian formalism developed in \cite{LV,GKL,beri0,CHY,GKR} (see also \cite{woo1,woo2}). Now we can make a more precise comparison. The extra coordinate of the $1+1$ formalism acts as a time translation in nonlocal trajectories, while in our case it is a variable independent of time. While our (regular) Hamiltonian system relies on a diffusion equation which is second-order in time derivatives and first-order in $r$ derivatives, the $1+1$ (constrained) analog is a chiral equation, i.e., first-order in both time and radial coordinate.


\section{The inverse problem}\label{inve}

To conclude, we classify the potentials admitted by exact solutions of the localized system. From Eqs.~\Eq{chiphi} and \Eq{locpheom}, one can write $\chi$ as a function of $\Phi$: $\chi[\Phi(r,x)]=V'[\Phi(r,x)]=\chi(r-2\gamma r,x)$. Three cases are possible.
\begin{enumerate}
\item $\p_r\Phi=0$. Then $V=\beta\Phi^2/2+{\rm const}$. This case is trivial as it corresponds to a local free theory.
\item $\p_r\Phi\neq 0$ and $\p_r\chi= 0$ (the potential does not depend on $r$). Then either (a) $\beta=0$ and $\B^2\Phi=0$, or (b) $\beta\neq 0$. An exact type (a) cosmological solution of the Klein--Gordon equation was given in \cite{cuta2}, while the homogeneous solution on Minkowski background ($\B=-\p_t^2$) is $\phi(r,t)=-2a_0r-6a_1rt+ \gamma a_0 t^2+\gamma a_1 t^3$ for $\chi(t)=-2a_0\gamma-6a_1\gamma t$ and some coefficients $a_0$, $a_1$. In case (b), the Minkowski homogeneous solution has $\chi=a_1 \cos (\sqrt{\beta} t)+a_2 \sin (\sqrt{\beta} t)$.
\item $\p_r\Phi\neq0\neq\p_r\chi$. $V'$ is an $r$-shifting function.
\end{enumerate}
To show that there exist nontrivial potentials of type (3), we specialize to a Minkowski background and homogeneous solutions. One can write the diffusion equation of $\chi[\Phi(r,x)]=\chi(r-2\gamma r,x)$ as
\be\label{use0}
u^2 \chi''-(2\gamma \dot u+\beta\Phi)\frac{\chi'}{1-2\gamma}+\beta\chi=0\,,
\ee
where $u\equiv \dot\phi$ and primes are derivatives with respect to $\Phi$. The values $\gamma=0$ and $\gamma=1/2$ are excluded because they correspond to case (1) and (2), respectively. If $\beta=0$, this equation can be written as a closed differential form, $\d \omega=0$, where $\omega=\ln \chi'-\gamma (1-2\gamma)^{-1} \ln u^2$. Therefore
\be\label{use}
\chi'= C u^{2\gamma/(1-2\gamma)}\,,
\ee
where $C$ is a constant. So far we have not made use of Eq.~\Eq{chiphi}, $\dot u=-\chi$. Differentiating Eq.~\Eq{use} with respect to time, one gets the final expression
\be\label{use2}
\frac{\chi'' (\chi')^{\frac1\gamma-3}}{\chi}=\frac{2\gamma C^{\frac1\gamma-2}}{2\gamma-1}={\rm const}.
\ee
Several potentials $V$ compatible with $\chi=V'$  solve this equation. For instance, $V\sim \Phi^n$ for $n=2(1-2\gamma)/(1-3\gamma)$, and one has $n>2$ if $0<\gamma<1/3$. This case includes the tachyon of exact bosonic ($n=3$) and approximate supersymmetric ($n=4$) cubic string field theory. In fact, the only modification of the action \Eq{locPh} is an extra mass term $-m^2\Phi(r,x)\Phi(r-2\gamma r,x)/2$, which changes Eq.~\Eq{chiphi} into $\chi(r,x) = (\B-m^2)\Phi(r,x)$, while Eqs.~\Eq{use0}--\Eq{use2} are unaffected.

Another example is $V\sim \e^{\lambda\Phi}$ ($\lambda$ generic, exponential or trigonometric potentials): then $\gamma =1/3$. Notice that, for the same value of $\gamma$, (infinitely) degenerate potentials such as trigonometric potentials are compatible with \Eq{use}, allowing the possibility to find also topologically nontrivial solutions.


\section{Conclusions}

The main advantages of the localized formulation are:
\begin{itemize}
\item The calculation of the energy-momentum tensor is much simpler than in the nonlocal case.
\item The diffusion equation is not only included automatically but is also needed for obtaining the same energy-momentum tensor. This supports the idea that all exact solutions of the original nonlocal model, if they exist, obey a diffusion equation. However, this evidence is not conclusive as it is not possible to check whether the exact solution of bosonic cubic SFT \cite{KORZ} satisfies a diffusion equation.
\item The diffusion equation helps in finding solutions of the system, both exact and approximate. We have given a recipe for the construction of nontrivial exact solutions, together with a few examples.
\item The Cauchy problem is well-defined, as there are only four initial conditions to be specified.
\item For the same reason, the Hamiltonian and conjugate momenta are easily constructed.
\item Quantization of the scalar fields forms a finite algebra.
\end{itemize}
On the other hand:
\begin{itemize}
\item There are models of physical interest which cannot be localized in the sense specified by Eqs.~\Eq{act}--\Eq{locch}. The supersymmetric string tachyon is one case, as mentioned in the introduction. However, one can study the solutions of the exact system via those of an approximate, localizable model \cite{roll}.
\item The diffusion equation method does not provide any existence conditions for the solutions, as the parameters $\gamma$ and $\beta$ are unconstrained. In \cite{roll,cuta2} we gave some examples of solutions with particular metrics and values of these parameters. Tachyon solutions of string field theory which obey the diffusion equation are actually approximated, although with high accuracy \cite{roll}.
\item The gravitational sector is treated as local.
\end{itemize}
The study of exact solutions for a given metric and potential, along the lines of Section \ref{inve}, will be important in the further understanding of the localized systems.


\ack

The work of G.C.\ is supported by a Marie Curie Intra-European Fellowship under contract MEIF-CT-2006-024523. The work of M.M.\ and G.N.\ is partly supported by INFN of Italy.


\end{document}